\documentclass[a4paper]{jpconf}
\usepackage{amsmath}
\usepackage{graphicx}
\usepackage{bm} 
\usepackage{tikz}
\usetikzlibrary{snakes}

\allowdisplaybreaks[1]
\def\be#1\ee{\begin{equation}#1\end{equation}}
\def\bal#1\eal{\begin{align}#1\end{align}}
\def\bat#1\eat{\begin{alignat}{2}#1\end{alignat}}
\def\bmu#1\emu{\begin{multline}#1\end{multline}}
\def\bga#1\ega{\begin{gather}#1\end{gather}}
\newcommand{\ba}{\begin{array}}
\newcommand{\ea}{\end{array}}
\newcommand{\n}{\notag}

\renewcommand{\d}{\partial}

\renewcommand{\cal}{\mathcal}

\newcommand{\scs}{\scriptstyle}

\begin{document}

\title{The infrared fixed point of Landau gauge Yang-Mills theory: %
A renormalization group analysis}

\author{Axel Weber}

\address{Instituto de F\'isica y Matem\'aticas,
Universidad Michoacana de San Nicol\'as de Hidalgo,
Edificio C-3, Ciudad Universitaria, A. Postal 2-82, 58040 Morelia, 
Michoac\'an, Mexico}

\ead{axel@ifm.umich.mx}

\begin{abstract}
The infrared behavior of gluon and ghost propagators in Landau
gauge Yang-Mills theory has been at the center of an intense debate over
the last decade. Different solutions of the Dyson-Schwinger equations show
a different behavior of the propagators in the infrared: in the so-called 
scaling solutions both propagators follow a power law, while in the
decoupling solutions the gluon propagator shows a massive behavior.
The latest lattice results favor the decoupling solutions. In this 
contribution, after giving a brief overview of the present status of
analytical and semi-analytical approaches to the infrared regime of
Landau gauge Yang-Mills theory, we will show how Callan-Symanzik 
renormalization group equations in an epsilon expansion reproduce 
both types of solutions and single out the decoupling solutions as the 
infrared-stable ones for space-time dimensions greater than two, in 
agreement with the lattice calculations.
\end{abstract}

\section{Introduction}

The infrared regime of quantum chromodynamics is one of the prime examples
of non-perturbative physics giving rise to such fascinating phenomena as
the confinement of quarks and the dynamical breaking of chiral symmetry. 
The only method so far to obtain 
quantitatively reliable results directly from the underlying theory has 
been lattice gauge theory, and it has done so only in a limited fashion.
For a long time, it was generally believed that analytical tools could not
be of much help in the quantitative description of the aforementioned
phenomena. This point of view is now slowly changing. In this contribution,
before describing our own approach to the subject to which the title refers, 
we will try to give an overview, from our perspective, of what has been 
achieved so far in (the pure gauge sector of) infrared quantum chromodynamics 
with analytical and semi-analytical methods. Needless to say,
progress with these methods has also been limited, but there have been 
some recent very promising developments.

The story begins in 1978 with the publication of Gribov's seminal work on
the quantization of non-abelian gauge theories \cite{Gri78}. It was clear
by that time that for the quantization of a gauge theory in the continuum 
the gauge needs to be fixed, and that, at least for perturbation theory, 
covariant gauges are particularly convenient. Specializing to the Landau gauge,
the functional integral over the gauge field then includes two characteristic
pieces: a functional delta distribution that implements the gauge condition
$\d_\mu A^a_\mu = 0$, and a functional determinant (Jacobian) that arises
when one changes the integration variables from the general gauge fields 
$A^a_\mu$ to gauge equivalent fields $\bar{A}^a_\mu$ that fulfill the gauge 
condition, and the corresponding gauge transformations $U$, 
$A^a_\mu = (\bar{A}^U)^a_\mu$. The delta distribution can conveniently be 
written with the help of a bosonic auxiliary field called the Nakanishi-Lautrup
field,%\footnote{The author is grateful to David Dudal for pointing out to
%him the convenience of the formulation with the Nakanishi-Lautrup field as
%opposed to the representation of the delta distribution as a limit of
%Gaussians with the gauge fixing parameter $\xi$ tending to zero.}
\be
\delta (\d_\mu A^a_\mu) = \int D [B] \, \exp \left(
-i \int d^D x \, B^a \d_\mu A^a_\mu \right) \,, \label{deltaf}
\ee
while the Jacobian is represented in a local way as a functional integral
over (fermionic scalar) ghost and antighost fields,
\be
\det (-\d_\mu D^{ab}_\mu) = \int D [c, \bar{c}] \, \exp \left(
\int d^D x \, \bar{c}^a \d_\mu D^{ab}_\mu c^b \right) \,. \label{fpdet}
\ee
We are generally working in $D$-dimensional Euclidean space-time, and
the covariant derivative (in the adjoint representation) is defined as
\be
D^{ab}_\mu = \delta^{ab} \d_\mu - g f^{abc} A^c_\mu \,.
\ee
In the functional integral of the theory (for gauge invariant operators)
the integral over the gauge transformations $U$ then factorizes and can
consistently be omitted, and one is left in the pure gauge sector with
an integral over $A^a_\mu$, $B^a$, $c^a$ and $\bar{c}^a$, and the
(negative of the) Faddeev-Popov action
\be
S_{\text{FP}} = \int d^D x \left( \frac{1}{4} F^a_{\mu\nu} F^a_{\mu\nu} 
+ i B^a \d_\mu A^a_\mu + \d_\mu \bar{c}^a D^{ab}_\mu c^b \right)
\label{fadpop}
\ee
in the exponent.

The original gauge symmetry of the Yang-Mills theory is manifestly broken
by the gauge fixing procedure, however, the integration measure and the
Faddeev-Popov action are invariant under the global fermionic
BRST transformation $s$ defined as
\bat
s A^a_\mu &= D^{ab}_\mu c^b \,, &\qquad s B^a &= 0 \,, \n \\
s c^a &= \frac{1}{2} g f^{abc} c^b c^c \,, &\qquad s \bar{c}^a &= i B^a \,.
\label{BRST}
\eat
An important property of this transformation (including the transformation
of the matter fields) is its nilpotency, $s^2 = 0$.
Note that the transformation of the gluon field has the form of an 
infinitesimal gauge transformation with the ghost field replacing the
generator of the gauge transformation. The Ward identities corresponding to 
the BRST symmetry are the Slavnov-Taylor identities for the correlation 
functions.

As it turns out, the beta function for this theory is negative giving rise
to the celebrated asymptotic freedom of non-abelian gauge theories and the 
applicability of perturbation theory in the ultraviolet regime. 
In the language of renormalization group theory, 
non-abelian gauge theories have a trivial fixed
point (with vanishing renormalized coupling constant $g_R (\mu)$) which is
ultraviolet-attractive. On the other hand, the renormalized coupling constant 
runs into a Landau pole at the characteristic scale $\Lambda_{\text{QCD}}$,
and perturbation theory (around the trivial fixed point) becomes useless
in the infrared, i.e., for energy-momentum scales of the order of and below
$\Lambda_{\text{QCD}}$.

\section{Gauge copies}

In 1978, Gribov \cite{Gri78} observed that the gauge fixing procedure leading 
to the Faddeev-Popov action as described above is actually not consistent
in non-abelian theories. The problem arises in the change of variables
in the functional integral over the gauge field: the choice of a gauge-fixed
field $\bar{A}^a_\mu$, $\d_\mu \bar{A}^a_\mu = 0$, that lies on the same gauge
orbit as the initial field $A^a_\mu$, $A^a_\mu = (\bar{A}^U)^a_\mu$ for an
appropriate gauge transformation $U$, is not unique. In other words,
there exist different gauge fields, called gauge copies or Gribov copies, 
that fulfill the gauge condition and are, nevertheless, connected by a gauge 
transformation. In geometric language, the gauge orbits intersect the 
hypersurface defined by $\d_\mu \bar{A}^a_\mu = 0$ more than once.

In particular, if we look at an infinitesimal gauge transformation generated
by $\omega^a$, we have
\be
\d_\mu (A^\omega)^a_\mu = \d_\mu A^a_\mu + \d_\mu D^{ab}_\mu \omega^b \,.
\label{infgaugecopies}
\ee
Then it is possible to have both, $A^a_\mu$ and $(A^\omega)^a_\mu$,
fulfill the gauge condition if $\omega^a$ is a zero mode of the Faddeev-Popov
operator $(-\d_\mu D^{ab}_\mu)$. Hence Gribov's proposal to restrict the
integration over the gauge field to the region where the Faddeev-Popov
operator is positive definite, now called the (first) Gribov region and
denoted by $\Omega$. Note that at the boundary of the Gribov region, 
the (first) Gribov horizon $\d \Omega$, we have that 
$\det (-\d_\mu D^{ab}_\mu) = 0$ because the Faddeev-Popov operator develops 
a zero mode there.

Although gauge copies which are related by an infinitesimal gauge
transformation are excluded by the restriction to $\Omega$, it is not clear
that finitely related gauge copies cannot exist within $\Omega$. In fact,
it is now known that such copies do exist in $\Omega$, and the consequent
thing to do would be to restrict the integration over the gauge fields
further to a smaller region that is called the fundamental modular region
$\Lambda$. However, the restriction to $\Lambda$ is very hard to implement,
in numerical calculations (lattice gauge theory with Landau gauge fixing)
and even more so in analytical (or semi-analytical) calculations.
Fortunately, according to an argument by Zwanziger \cite{Zwa04},
the difference between the restrictions to $\Omega$ or to $\Lambda$ has no
effect on the correlation functions.

For the rest of this contribution, we will hence consider the restriction
of the integral over the gauge field to the Gribov region $\Omega$. One
of the important implications of this restriction as compared to an
unrestricted integral over the gauge field, according to an argument by
Gribov \cite{Gri78}, is the impossibility of a Landau pole for the renormalized
coupling constant to occur (at a momentum scale different from zero).
Realizing the restriction to $\Omega$ in an approximate way, Gribov was
furthermore able to predict the infrared behavior of the gluon and ghost
propagators. Here we will present an exact way to implement the restriction
of the integral over the gauge fields which is due to Zwanziger \cite{Zwa89} 
and based on Gribov's ideas, and come back to the propagators later.

Zwanziger found a way to realize the (Heavyside) theta functional for the
restriction of the integral over the gluon field to the Gribov region as a
Gaussian (in the thermodynamic limit):
\be
\int_\Omega D [A] = \int D [A] \, \theta (-\d_\mu D^{ab}_\mu)
= \int D [A] \, \exp \left( - \gamma^4 \int d^D x \, h(x) \right) \,,
\label{DArestrict}
\ee
with the horizon function $h(x)$ defined as
\be
h(x) = g^2 f^{abc} f^{cde} A^b_\nu \, 
\big[ (-\d_\mu D_\mu)^{-1} \big]^{ad} A^e_\nu \,.
\ee
For the correct normalization, the horizon condition
\be
\langle h(x) \rangle = 4 (N^2 - 1) \label{horizon}
\ee
has to be fulfilled (in an SU$(N)$ gauge theory). This condition fixes the 
new parameter $\gamma$ with the dimension of mass (in terms of 
$\Lambda_{\text{QCD}}$, or vice versa).

Now $h(x)$ is a complicated nonlocal functional because it contains
the inverse of the Faddeev-Popov operator. A local form of the theory, 
necessary for
the application of the usual quantum field theoretical machinery, can be 
achieved by introducing two additional auxiliary vector fields (and the
corresponding antifields), one of them bosonic $\phi^{ab}_\mu$ and the
other fermionic $\omega^{ab}_\mu$, each carrying two color indices (in
the adjoint representation). The gauge sector of the theory including the
restriction to the first Gribov region is then described by
a functional integral over $A^a_\mu$, $B^a$, $c^a$, $\bar{c}^a$, 
$\phi^{ab}_\mu$, $\bar{\phi}^{ab}_\mu$, $\omega^{ab}_\mu$ and
$\bar{\omega}^{ab}_\mu$, and the exponential of the (negative of the)
Gribov-Zwanziger action \cite{Zwa93}
\bal
S_{\text{GZ}} &= S_{\text{FP}} - \int d^D x \left(
\d_\mu \bar{\phi}^{ac}_\nu D^{ab}_\mu \phi^{bc}_\nu
- \d_\mu \bar{\omega}^{ac}_\nu D^{ab}_\mu \omega^{bc}_\nu
+ g f^{abc} \d_\mu \bar{\omega}^{ad}_\nu D^{be}_\mu c^e \phi^{cd}_\nu 
\right) \n \\
%&\phantom{= \cal{L}_{FP}} 
&\phantom{=} {}- \gamma^2 \int d^D x \left( g f^{abc} A^a_\mu \phi^{bc}_\mu
+ g f^{abc} A^a_\mu \bar{\phi}^{bc}_\mu + 4 (N^2 - 1) \gamma^2 \right) \,,
\label{grizwa}
\eal
see eq.\ \eqref{fadpop}. The horizon condition can be rewritten in terms
of the auxiliary fields as
\be
\big\langle g f^{abc} A^a_\mu 
(\phi^{bc}_\mu + \bar{\phi}^{bc}_\mu) \big\rangle 
= 8 (N^2 - 1) \gamma^2 \,. \label{horcond}
\ee

The Gribov-Zwanziger action leads to the tree-level gluon propagator
(in momentum space)
\be
\langle A^a_\mu (p) A^b_\nu (-q) \rangle 
= \frac{p^2}{p^4 + \lambda^4} \left( \delta_{\mu\nu}
- \frac{p_\mu p_\nu}{p^2} \right) \delta^{ab} (2 \pi)^D \delta (p-q) \,, 
\label{gluonGZ}
\ee
with the characteristic scale $\lambda$ defined as
\be
\lambda^4 = 2 g^2 N \gamma^4 \,. \label{lambda4}
\ee
For $p^2 \gg \lambda^2$, one recovers the usual (tree-level) ultraviolet
behavior, while the propagator is heavily suppressed for $p^2 \ll \lambda^2$
and even tends to zero in the limit $p^2 \to 0$. The tree-level ghost 
propagator, on the other hand, behaves like
\be
\langle c^a (p) \bar{c}^b (-q) \rangle \propto \frac{1}{p^4} \, \delta^{ab}
(2 \pi)^D \delta (p-q) \label{ghostGZ}
\ee
in the infrared regime $p^2 \ll \lambda^2$. The divergence in the limit 
$p^2 \to 0$ is much stronger than for the tree-level ghost propagator in
Faddeev-Popov theory that does not take the restriction to the Gribov region 
into account. It is clear, then, that the effect of the restriction to the
Gribov region is rather drastic for the infrared behavior of the propagators.
The forms \eqref{gluonGZ} and \eqref{ghostGZ} of the propagators 
have already been found by Gribov in his approximate description \cite{Gri78}.

\section{Dyson-Schwinger equations}

In an initially independent development, the Dyson-Schwinger equations
of quantum chromodynamics were approximately solved in the infrared regime.
As in the previous section, we here focus on the pure gauge sector, i.e., 
Yang-Mills theory. The Dyson-Schwinger equations are derived from the 
Faddeev-Popov action \eqref{fadpop}, and the restriction to the Gribov region 
is not implemented in any way. While this is clear from the independence of 
this development, the reader is justified to wonder why it
would make any sense to present the solution of Dyson-Schwinger equations
that ignore the restriction to the Gribov region, when we have argued at the
end of the last section that the restriction to the Gribov region has a 
severe effect on the infrared behavior of the propagators.

However, as Zwanziger has clarified \cite{Zwa02}, the restriction to the 
Gribov region actually leaves the Dyson-Schwinger equations unchanged:
Dyson-Schwinger equations are derived from the fact that the functional
integral over a (functional) derivative vanishes. When one restricts the 
functional integral over the gauge field in the Faddeev-Popov formulation 
to the Gribov region $\Omega$, contributions from the boundary 
$\d \Omega$ are to be expected.
However, there are no such contributions in this case since
the integrand vanishes on $\d \Omega$ due to the presence of the 
Faddeev-Popov determinant as remarked earlier, and the resulting 
Dyson-Schwinger equations are the same as for the unrestricted functional 
integral.

Solutions to the Dyson-Schwinger equations for the gluon and ghost
propagators have been found, in a certain approximation, first in
refs.\ \cite{SHA97,SHA98}. The results are an infrared suppressed gluon
propagator and an infrared enhanced ghost propagator as compared to the
tree-level propagators of Faddeev-Popov theory, which is qualitatively
consistent with the results \eqref{gluonGZ} and \eqref{ghostGZ} of the 
Gribov-Zwanziger framework. Then, with hindsight, we can simplify the
Dyson-Schwinger equations in the infrared regime by omitting the diagrams
that contain gluon loops, or in general, in the Dyson-Schwinger equation 
for a given correlation function only keep the diagrams with the biggest
number of ghost loops since the latter dominate the infrared behavior.
The dominance of the diagrams with the biggest number of internal ghost 
propagators in the infrared regime is sometimes referred to as ghost dominance.

Furthermore, Taylor's non-renormalization theorem \cite{Tay71} allows
us to replace the full ghost-gluon vertex by the bare one. That this
replacement is valid to a good approximation even in the non-perturbative
regime has been checked in lattice calculations \cite{CMM04}. With these
simplifications, the Dyson-Schwinger equations for the gluon and ghost
propagators take the form depicted in fig.\ \ref{IRDS}.
\begin{figure}
\bal
  \Big( \raisebox{-4pt}{\parbox{1,4cm}{\begin{center}
  \begin{tikzpicture}[scale=1.4]
   \begin{scope}[snake=coil, segment length=3pt, segment amplitude=2pt,%
   line before snake=1.5pt]
   \draw[snake] (0.2,0) -- (1,0) node[below] {$\scs p$};
   \end{scope}
   \fill (0.58,0) circle (1.75pt);
  \end{tikzpicture}
  \end{center}}} \Big)^{-1}
  &= \, Z_A (p^2 \delta_{\mu\nu} - p_\mu p_\nu) \,
  - \parbox{3.1cm}{\begin{center}
  \begin{tikzpicture}[>=stealth,scale=1.4]
   \begin{scope}[snake=coil, segment length=3pt, segment amplitude=2pt,%
   line before snake=1.5pt]
   \draw[snake] (-0.45,0) -- (0.05,0);
   \draw[snake] (0.98,0) -- (1.44,0) node[below] {$\scs p$};
   \end{scope}
   \draw[dash pattern=on 1.5pt off 1.2pt] (0.5,0) circle (0.5cm);
   \draw[->,very thin] (0.85,-0.35) -- (0.88,-0.32);
   \draw[->,very thin] (0.15,-0.35) -- (0.18,-0.38);
   \draw[->,very thin] (0.85,0.35) -- (0.82,0.38);
   \draw[->,very thin] (0.15,0.35) -- (0.12,0.32);
   \fill (0.5,0.5) circle (1.6pt);
   \fill (0.5,-0.5) circle (1.6pt);
   \fill (0,0) circle (1pt);
   \fill (1,0) circle (1pt);
  \end{tikzpicture}
  \end{center}} \n
\\
  \Big( \raisebox{-4pt}{\parbox{1.4cm}{\begin{center}
  \begin{tikzpicture}[>=stealth,scale=1.4]
   \draw[dash pattern=on 1.5pt off 1.5pt] (0.2,0) -- (1,0) %
node[below] {$\scs p$};
  \fill (0.6,0) circle (1.7pt);
   \draw[->,very thin] (0.33,0) -- (0.3,0);
   \draw[->,very thin] (0.83,0) -- (0.8,0);
  \end{tikzpicture}
  \end{center}}} \Big)^{-1}
  &= \, Z_c p^2 \,
  - \parbox{3.1cm}{\begin{center}
  \begin{tikzpicture}[>=stealth,scale=1.4]
   \begin{scope}[snake=coil, segment length=3pt, segment amplitude=2pt]
   \foreach \x in {-5,5,...,165}
     \draw[snake] [xshift=0.5cm] (\x+19:0.5cm) -- (\x:0.5cm);
   \end{scope}
   \draw[dash pattern=on 1.5pt off 1.5pt] (-0.4,0) -- (1.4,0) %
node[below] {$\scs p$};
   \draw[->,very thin] (1.18,0) -- (1.15,0);
   \draw[->,very thin] (0.77,0) -- (0.74,0);
   \draw[->,very thin] (0.27,0) -- (0.24,0);
   \draw[->,very thin] (-0.22,0) -- (-0.25,0);
   \fill (0.5,0) circle (1.6pt);
   \fill (0.5,0.5) circle (1.65pt);
   \fill (0,0) circle (1pt);
   \fill (1,0) circle (1pt);
  \end{tikzpicture} 
  \end{center}} \n
\eal
\caption{\label{IRDS} Simplified Dyson-Schwinger equations for the gluon and
ghost propagators containing the dominant contributions in the infrared
regime.}
\end{figure}
This simplified system of equations not only correctly describes the extreme
infrared limit of the theory, but it also gives a qualitative description
of the extreme ultraviolet, asymptotically free regime, and it can thus
serve as a useful laboratory for the study of several properties of the 
complete theory \cite{Dal11}. For a quantitatively correct description of
the ultraviolet regime, in particular the correct value of the (one-loop)
beta function, however, one needs to take additional contributions into
account in the first equation of fig.\ \ref{IRDS}, and more sophisticated
approximation schemes have been devised to this end \cite{FA02,FMP09}.

If one is exclusively interested in the extreme infrared regime, the 
simplified Dyson-Schwinger equations can even be solved analytically.
In order to describe the solutions, we will introduce some notation:
we write the propagators in terms of their dressing functions 
$G(p^2)$ and $F(p^2)$,
\bal
\langle A^a_\mu (p) A^b_\nu (-q) \rangle 
&= \frac{G(p^2)}{p^2} \left( \delta_{\mu\nu} - \frac{p_\mu p_\nu}{p^2} \right) 
\delta^{ab} (2 \pi)^D \delta (p-q) \,, \n \\
\langle c^a (p) \bar{c}^b (-q) \rangle 
&= \frac{F(p^2)}{p^2} \, \delta^{ab} (2 \pi)^D \delta (p-q) \,.
\eal
For the analytical solutions, we make power-like ansaetze for the dressing
functions (this behavior is familiar from the theory of critical 
phenomena):
\be
G(p^2) \propto (p^2)^{-\alpha_G} \,, \qquad
F(p^2) \propto (p^2)^{-\alpha_F} \,. \label{powerans}
\ee

Inserting the ansaetze \eqref{powerans} in the system of approximate
Dyson-Schwinger equations in fig.\ \ref{IRDS}, one obtains, for any dimension
$D$ between two and four, \emph{two} consistent solutions known as
scaling solutions \cite{Zwa02,LS02}. One of these solutions is
\be
\alpha_F (D) = \frac{D-2}{2} \,, 
\qquad \alpha_G (D) = - \frac{D}{2} \label{DSsol1}
\ee
in dimension $D$, for $2 < D < 4$ (this solution is not entirely consistent
in dimensions two and four due to logarithmic corrections). The infrared
exponents of the other scaling solution have to be obtained, for general
$D$, from the solution of a transcendental equation. A good approximation
(within 2\% error for any $D$ with $2 \le D \le 4$) is given, however, by 
the simple formulae
\be
\alpha_F (D) = \frac{D-1}{5} \,, 
\qquad \alpha_G (D) = - \frac{16 - D}{10} \,. \label{DSsol2}
\ee

For both solutions (also the exact version of the second solution), the 
infrared exponents fulfill the sum rule
\be
\alpha_G + 2 \alpha_F = \frac{D-4}{2} \label{sumrule}
\ee
in any dimension $D$. The solutions also have in common that $\alpha_F > 0$
and $1 + \alpha_G < 0$ (for $D>2$ in the first solution), implying that
the ghost propagator diverges more strongly in the infrared than the 
(Faddeev-Popov) tree-level propagator, while the gluon propagator is
strongly suppressed in the infrared, i.e., it tends to zero in the limit
$p^2 \to 0$, just like the tree-level propagators in the Gribov-Zwanziger
framework. 

Taking into account that the Dyson-Schwinger equations are not altered 
by the restriction of the functional integral to the Gribov region 
as argued at the beginning of this section, we interpret the solutions 
\eqref{DSsol1} and \eqref{DSsol2} as (approximately) incorporating the
quantum corrections to the tree-level results \eqref{gluonGZ} and 
\eqref{ghostGZ} presented in the previous section. It is also clear
that the power-like behavior \eqref{powerans} of the solutions can only
be obtained by summing up the quantum corrections (in some approximation) 
to \emph{any} loop order. On the other hand, it is a priori not so clear
which of the two solutions \eqref{DSsol1} and \eqref{DSsol2} is the 
physically relevant one for general dimension $D$. Also, one would like to 
have a systematic scheme to improve on the approximations employed in the
derivation of the Dyson-Schwinger equations in fig.\ \ref{IRDS},
and it has turned out to be very difficult to formulate such a scheme.

An important consequence of Taylor's non-renormalization theorem is the
natural definition of a running coupling constant as
\be
g_R^2 (p^2) = (p^2)^{(D-4)/2} \, G (p^2) \, F^2 (p^2) \, g^2 \,.
\label{runncoupl}
\ee
The factor $(p^2)^{(D-4)/2}$ compensates for the dimension of the bare
coupling constant $g^2$ (the dressing functions $G(p^2)$ and $F(p^2)$ are 
by definition dimensionless), so that $g_R^2 (p^2)$ is a dimensionless 
quantity. Then it is clear from the sum rule \eqref{sumrule} that the
running coupling constant tends to a finite fixed point value in the infrared
limit. This value can be determined analytically for the two scaling
solutions \eqref{DSsol1} and \eqref{DSsol2}, and it turns out to vary 
continuously with the dimension $D$. As an example, we give the (exact) 
fixed point value in $D=4$ for the second solution \eqref{DSsol2},
\be
\frac{N g_R^2}{4 \pi} = 8{.}915 \ldots \,,
\ee
and mention that $g_R^2$ tends to zero in the limit $D \to 2$ for the
first solution \eqref{DSsol1}.

Finally, we mention that the same results have also been obtained (in $D=4$
dimensions) with a different technique, the functional renormalization group 
equations \cite{FMP09,PLN04} which, however, have many features in common 
with the Dyson-Schwinger equations.

\section{Decoupling solutions}

One of the characteristic features of the scaling solutions (including their
tree-level forms \eqref{gluonGZ}, \eqref{ghostGZ}) is the infrared
enhancement of the ghost propagator. In fact, it may be shown
\cite{Zwa93} that the horizon condition \eqref{horizon} is equivalent to
\be
\lim_{p^2 \to 0} F^{-1} (p^2) = 0 \,, \label{infdivgh}
\ee
the infrared divergence of the ghost dressing function. 
It is well understood by now that implementing the condition \eqref{infdivgh}
necessarily leads to the scaling solutions, be it in the tree-level form
\eqref{gluonGZ} and \eqref{ghostGZ} or in the ``resummed'' form
\eqref{DSsol1} or \eqref{DSsol2}.

There is another important aspect of the Gribov-Zwanziger framework
\cite{Zwa94}: the horizon condition is identical to the Kugo-Ojima 
confinement criterion \cite{KO79,Kug95} which guarantees (in Faddeev-Popov 
theory) the existence of a global color charge that is BRST-exact and 
hence implies confinement in the sense that all physical states are color 
singlets. Kugo and Ojima construct the physical Hilbert space of non-abelian
gauge theories starting from BRST invariance and show the unitarity of the 
$S$-matrix \cite{KO79}. The equivalence of the confinement criterion
(and hence the horizon condition) to the infrared divergence
\eqref{infdivgh} of the ghost dressing function in the Landau gauge
has been shown in this framework, too \cite{Kug95}.

At first sight, everything seems to fit together nicely: the restriction of 
the functional integral over the gluon field to the Gribov region, necessary 
because of the existence of gauge copies in non-abelian gauge theories, 
leads to the horizon condition or the Kugo-Ojima criterion and hence to the 
confinement of color charges. However, there is a fundamental flaw in this 
argument: the BRST symmetry of the Faddeev-Popov action can be extended to 
the Gribov-Zwanziger action \eqref{grizwa} only for $\gamma^2 = 0$,
in which case the integration over the gluon field is effectively 
unrestricted, see eq.\ \eqref{DArestrict}. In other words, the 
restriction to the Gribov region breaks the BRST invariance on which the 
whole Kugo-Ojima construction rests.

There is a simple intuitive argument that shows how the breakdown of BRST
invariance is an inevitable consequence of the restriction to the Gribov
region \cite{DGS08}: the BRST transformation of the gluon field has the 
form of an infinitesimal gauge transformation, as already remarked following
eq.\ \eqref{BRST}. But close to the Gribov horizon, an infinitesimal
gauge transformation can relate gauge copies, see eq.\ 
\eqref{infgaugecopies}, and hence the transformed field falls outside 
the Gribov region.

The violation of BRST symmetry is not in itself a catastrophe for the
theory. Many of us have grown up thinking of gauge invariance and BRST
invariance as synonyms, mostly because the (global) BRST symmetry implies
the Slavnov-Taylor identities. However, this is really only true for the
Faddeev-Popov action and an unrestricted functional integral over the
gluon field. If we restrict the intregration to the Gribov region, as it
turns out to be necessary in order to avoid gauge copies, the (gauge
fixed) quantum theory is not invariant under BRST transformations
any more. Of course, this is a serious problem for the construction of the
Hilbert space of the theory, just because the standard construction (of
Kugo and Ojima) is based on BRST symmetry. A solution to this problem has 
not yet been devised.

But there is yet another twist to our story. With the analytical and
semi-analytical predictions of the propagators in the infrared regime at
hand, it seemed natural to corroborate these predictions with the help
of completely non-perturbative lattice calculations. While it is not
necessary (nor natural) to fix the gauge in these calculations, it is also
not too difficult to implement the Landau gauge on the lattice: for a
given gauge field configuration $A^a_\mu$, consider the functional
\be
\int d^D x \, (A^U)^a_\mu (A^U)^a_\mu \label{gaugefunct}
\ee
for all possible gauge transformations $U$, i.e., along the gauge orbit
that contains $A^a_\mu$. Then the extrema (or saddle points) of this
functional along the orbit correspond to the fields $\bar{A}^a_\mu = 
(A^U)^a_\mu$ that fulfill the gauge condition $\d_\mu \bar{A}^a_\mu = 0$,
the local minima correspond to fields $\bar{A}^a_\mu$ in the first Gribov
region $\Omega$, and the absolute minima to fields in the fundamental
modular region $\Lambda$. For the huge lattices that are necessary to
measure the infrared behavior of the propagators, it is extremely difficult
(numerically costly) to determine the absolute minimum of the functional
\eqref{gaugefunct} along a given gauge orbit, however, it is relatively easy 
to find a local minimum and hence to restrict the gauge fields to the 
Gribov region.

The first lattice results for the propagators in the Landau gauge 
(with the restriction to the Gribov region $\Omega$) show
a clear increase of the ghost dressing function $F(p^2)$ towards the
infrared, and a maximum and subsequent marked decrease of the gluon 
dressing function $G(p^2)$ as one follows $p^2$ from the ultraviolet to 
the infrared, in agreement with the scaling solutions of the Dyson-Schwinger
equations. However, it was difficult to discern for a long time whether 
the ghost dressing function would actually diverge in the limit $p^2 \to 0$ 
and whether the gluon propagator itself would vanish in the same limit.

Meanwhile, it was discovered that the Dyson-Schwinger equations of the
theory permit another type of solutions now known as decoupling solutions
\cite{AN04,BBL06,Fra08} (see also \cite{FMP09}). 
In these solutions, the ghost propagator is infrared enhanced only by a 
finite constant, i.e., the ghost dressing functions tends toward a finite 
constant in the limit $p^2 \to 0$, while the gluon 
propagator has a finite (nonzero) limiting value similarly to the propagator 
of a massive particle. Then one still has ghost dominance
in the sense that the gluon propagator is infrared suppressed in comparison
with the ghost propagator, and the reduction of the full Dyson-Schwinger
equations to the approximation represented in fig.\ \ref{IRDS} is
justified in the infrared regime.

In terms of the infrared exponents defined in eq.\ \eqref{powerans}, one
has for the decoupling solutions that
\be
\alpha_F = 0 \,, \qquad \alpha_G = -1 
\ee
in any dimension $D$ (coincident with the first scaling solution 
\eqref{DSsol1} in the limit $D \to 2$). In particular, the decoupling 
solutions do not respect the sum rule \eqref{sumrule} for the exponents 
(except at $D=2$). Furthermore, the running coupling 
constant defined in eq.\ \eqref{runncoupl} tends to zero in the limit 
$p^2 \to 0$ for the decoupling solutions (for dimensions $D > 2$).

The discovery of the decoupling solutions of the Dyson-Schwinger equations
initiated a long and often passionate debate in the community about 
the true infrared behavior of the Landau gauge propagators. In 2007, at 
the lattice conference in Regensburg, three lattice groups
presented their results on the infrared behavior of the propagators in
Landau gauge Yang-Mills theory on very large lattices 
\cite{BIM07,CM07,SSL07}. The majority of the community has interpreted
these results as confirming the \emph{decoupling solutions} of the
Dyson-Schwinger equations in three and four space-time dimensions. In the 
same year, lattice calculations in \emph{two} dimensions \cite{Maa07} 
confirmed the realization of the second scaling solution \eqref{DSsol2}. 

At first sight,
the decoupling solutions are at odds with the tree-level propagators
\eqref{gluonGZ} and \eqref{ghostGZ} and the horizon condition in the form
\eqref{infdivgh} that result in the Gribov-Zwanziger 
scenario. However, indications were found in ref.\ \cite{DGS08} (see also 
\cite{DSV11}) that the auxiliary fields form a condensate 
$\langle \bar{\phi}^{ab}_\mu \phi^{ab}_\mu \rangle$ (among other possible 
condensates). As a consequence, the horizon condition \eqref{horcond} does
\emph{not} imply the infrared divergence \eqref{infdivgh} of the ghost 
dressing function any more, and through the linear coupling of the fields 
$\phi^{ab}_\mu$ and $\bar{\phi}^{ab}_\mu$ to the gluon field, see eq.\ 
\eqref{grizwa}, the condensate leads to a mass term for the gluons.

In the simplest case \cite{DGS08} in this so-called refined Gribov-Zwanziger
framework, i.e., with the condensates taken into account, the tree-level 
propagators are of the form
\be
\langle A^a_\mu (p) A^b_\nu (-q) \rangle 
= \frac{p^2 + M^2}{p^4 + M^2 p^2 + \lambda^4} \left( \delta_{\mu\nu}
- \frac{p_\mu p_\nu}{p^2} \right) \delta^{ab} (2 \pi)^D \delta (p-q) 
\label{gluonRGZ}
\ee
(see eq.\ \eqref{lambda4} for the definition of the characteristic scale 
$\lambda$), and
\be
\langle c^a (p) \bar{c}^b (-q) \rangle \propto \frac{1}{p^2} \, \delta^{ab}
(2 \pi)^D \delta (p-q) \label{ghostRGZ}
\ee
in the infrared regime. The gluon propagator can also
be written in terms of an effective momentum-dependent mass 
$M_{\text{eff}} (p^2)$,
\be
\frac{p^2 + M^2}{p^4 + M^2 p^2 + \lambda^4} = 
\frac{1}{p^2 + M_{\text{eff}}(p^2)} \,, \label{gluonRGZ2}
\ee
where
\be
M_{\text{eff}}(p^2) = \frac{\lambda^4}{p^2 + M^2} \,. \label{effmass}
\ee
In four space-time dimensions, a successful fit of
the tree-level gluon propagator in the refined Gribov-Zwanziger framework 
with an additional $\langle A^a_\mu A^a_\mu \rangle$-condensate,
\be
\langle A^a_\mu (p) A^b_\nu (-q) \rangle 
= \frac{p^2 + M^2}{p^4 + (M^2 + m^2) p^2 + \lambda^4 + M^2 m^2} 
\left( \delta_{\mu\nu} - \frac{p_\mu p_\nu}{p^2} \right) 
\delta^{ab} (2 \pi)^D \delta (p-q) \,, \label{gluonRGZ3}
\ee 
to the lattice data in the infrared and intermediate momentum regimes has 
been performed in ref.\ \cite{DOV10} (see also \cite{CDM11}). It is
important to clarify that the different mass parameters that have appeared 
so far are not additional parameters of the theory, but can rather be
determined, at least in principle, as multiples of $\Lambda_{\text{QCD}}$
from a calculation of the effective potential \cite{DGS08}.

At this point, from the perspective of the Dyson-Schwinger equations,
the decoupling solution as well as both of the scaling solutions are
consistent and valid solutions, and from these equations alone there is
no reason to prefer one of the solutions over the others. In the refined
Gribov-Zwanziger framework, on the other hand, the decoupling solution
with the tree-level propagators \eqref{gluonRGZ} or \eqref{gluonRGZ3}
and \eqref{ghostRGZ} is preferred according to the lowest-order 
calculation of the effective potential for
various local composite operators \cite{DSV11}. The
problem with this latter approach is rather of a technical nature: it 
appears extremely cumbersome to carry the calculations to higher orders.

We finally mention an important property of the scaling as well as the
decoupling solutions: all the solutions found so far violate positivity,
i.e., the spectral function in the K\"all\'en-Lehmann representation of
the gluon propagator is not positive semidefinite implying that the gluon 
can\emph{not} be a physical (massive or massless) particle (see, e.g., refs.\
\cite{SHA98,FMP09,DGS08,Zwa91,CMT05}).

\section{Callan-Symanzik equations}

\subsection{A gluon mass term}

For the rest of this contribution, we will describe a recent approach
to the subject from a renormalization group perspective \cite{Web11}.
We start from the Dyson-Schwinger equations of Faddeev-Popov theory
which, as explained before, do not change when one restricts the functional
integral over the gluon field to the Gribov region $\Omega$. Since the
iterative solution of the Dyson-Schwinger equations generates the
perturbative expansion, the same is true for the perturbative series.
On the other hand, the (re)normalization conditions that define the
quantum theory can certainly change as a result of the restriction.

One of the most important consequences of the restriction to the Gribov
region, as mentioned in the last section, is the breaking of BRST
invariance. The resulting violation of the Slavnov-Taylor identites leads,
almost inevitably, to the generation of a gluon mass term as the only
possible properly relevant term in the (effective) action with the field
content of Faddeev-Popov theory, i.e., without the additional auxiliary
fields of the Gribov-Zwanziger action. Note that a ghost mass term cannot be 
induced for the same kinematical reasons that lead to Taylor's
non-renormalization theorem. We will, hence, include a gluon mass term 
\be
\int d^D x \, \frac{1}{2} A^a_\mu m^2 A^a_\mu \label{mass}
\ee
in the Faddeev-Popov action.

Such an action has been considered first, as a special case among a much
more general class of actions, in ref.\ \cite{CF76} where the extension
of the BRST transformation and the Slavnov-Taylor identities to the
massive case is discussed (however, the Nakanishi-Lautrup field is not
introduced in ref.\ \cite{CF76}). The Faddeev-Popov action with a gluon
mass term has also been used in a recent perturbative calculation
\cite{TW10,TW11} to which we come back later. Since the mass term originates
from the breaking of the BRST symmetry as a consequence of the restriction to
the Gribov region, we expect the (renormalized) mass parameter to be of
the order of the characteristic scale $\Lambda_{\text{QCD}}$. It is then
clear that the introduction of the mass term has no important effect
in the ultraviolet regime, i.e., for momenta $p^2 \gg \Lambda^2_{\text{QCD}}$.
In particular, the value of the ultraviolet beta function is unchanged and,
of course, the theory is asymptotically free. This is consistent with the
fact that the restriction to the Gribov region is irrelevant for the
perturbative expansion in the ultraviolet regime where only ``small''
fields $A^a_\mu$ contribute.

Now, we would at least naively expect that in the extreme infrared regime 
$p^2 \ll m^2$ the mass term \eqref{mass}
dominates over the other term in the action that is quadratic in $A^a_\mu$,
\be
\int \frac{d^D p}{(2 \pi)^D} \, A^a_\mu (-p) \left( p^2 \delta_{\mu\nu}
- p_\mu p_\nu \right) A^a_\nu (p) \,. \label{cinetic}
\ee
In fact, this is true also
for the loop corrections as we will now illustrate with the help of a 
relevant example: consider the one-loop ghost self-energy
\be
\parbox{3.1cm}{\begin{center}
  \begin{tikzpicture}[>=stealth,scale=1.4]
   \begin{scope}[snake=coil, segment length=3pt, segment amplitude=2pt]
   \foreach \x in {-5,5,...,165}
     \draw[snake] [xshift=0.5cm] (\x+19:0.5cm) -- (\x:0.5cm);
   \end{scope}
   \draw[dash pattern=on 1.5pt off 1.2pt] (-0.4,0) -- (1.4,0) %
node[below] {$\scs p$};
   \draw[->,very thin] (1.18,0) -- (1.15,0);
   \draw[->,very thin] (0.51,0) -- (0.48,0) node[below] {$\scs p-k$};
   \draw[->,very thin] (-0.22,0) -- (-0.25,0);
   \draw[very thin] (0.5,0.5) -- (0.5,0.5) node[above] {$\scs k$};
   \fill (0,0) circle (1.2pt);
   \fill (1,0) circle (1.2pt);
  \end{tikzpicture} 
  \end{center}}
= N g^2 \delta^{ab} \int \frac{d^D k}{(2 \pi)^D} \, p_\mu \, 
\frac{1}{(p - k)^2} \, (p_\nu - k_\nu) \, \frac{1}{k^2 + m^2} 
\left( \delta_{\mu \nu} - \frac{k_\mu k_\nu}{k^2} \right) \,. \label{ghostSE}
\ee
It is worth mentioning that the tree-level gluon propagator in Landau gauge
is exactly transverse in momentum space even though the mass term \eqref{mass} 
has a longitudinal part. The diagram \eqref{ghostSE} can be evaluated for
arbitrary dimension $D$, e.g., by introducing Feynman parameters. We will
here concentrate on $D=2$, which will turn out to play a particularly 
important role, and comment on other dimensions later. Then, evaluating
\eqref{ghostSE} in $D=2$ and expanding the result in powers of $(p^2/m^2)$
for the extreme infrared behavior gives
\be
- \frac{1}{2} \, \frac{N g^2}{4 \pi} \, \delta^{ab} \, \frac{p^2}{m^2}
\left[ \ln \frac{p^2}{m^2} - 1 - \frac{1}{2} \, \frac{p^2}{m^2} 
+ \cal{O} \left( \big( p^2/m^2 \big)^2 \right) \right] \,. \label{ghostSE1}
\ee

On the other hand, neglecting the term \eqref{cinetic} in the action 
corresponds to the replacement
\be
\frac{1}{k^2 + m^2} \left( \delta_{\mu \nu} - \frac{k_\mu k_\nu}{k^2} 
\right) \longrightarrow
\frac{1}{m^2} \left( \delta_{\mu \nu} - \frac{k_\mu k_\nu}{k^2} \right)
\label{gluonreplace}
\ee
for the tree-level gluon propagator. Making this replacement in the loop
diagram \eqref{ghostSE} leads to the result
\be
- \frac{1}{2} \, \frac{N g^2}{4 \pi} \, \delta^{ab} \, \frac{p^2}{m^2}
\left( \frac{2}{\epsilon} + \gamma_E - \ln (4 \pi)
+ \ln \frac{p^2}{\kappa^2} \right) \label{ghostSE2}
\ee
(plus terms of order $\epsilon$) in $D = 2 + \epsilon$ dimensions, with an 
arbitrary unit of mass $\kappa$. The appearance of an ultraviolet divergency 
in two dimensions is due to the radical change in the ultraviolet behavior of 
the gluon propagator. To be precise, the symbol $g$ in eq.\ \eqref{ghostSE2} 
stands for the bare coupling constant multiplied with a power of $\kappa$ such 
that $g$ has the dimension of mass, and it coincides with the 
coupling constant appearing in eq.\ \eqref{ghostSE1} only in the limit
$D \to 2$.

The results \eqref{ghostSE1} and \eqref{ghostSE2} are obviously different.
However, if we implement an appropriate normalization condition for the term 
proportional to $p^2$ in the proper ghost two-point function, then \emph{after 
renormalization} the two results will only differ by contributions
of the order $(p^2/m^2)$ relative to the leading term, since renormalization
takes care of the constants multiplying $p^2$. Expressions \eqref{ghostSE1} 
and \eqref{ghostSE2} are then equivalent in the infrared limit, and in this
sense the replacement \eqref{gluonreplace} of the tree-level gluon propagator 
is justified. The argument can easily be extended to any dimension $D$ in
the range $2 \le D \le 4$.

We can, hence, neglect the term \eqref{cinetic} in the action as long as
we are only interested in describing the extreme infrared regime of the
theory. These modifications, the introduction of a mass term for the gluons
and the subsequent omission of the term \eqref{cinetic}, have profound
consequences for the renormalization group anaylisis of the theory. We
begin this analysis by considering the free ($g=0$) part of the action
which is now given by
\be
S^0_{\text{IR}} = \int d^D x \left( \frac{1}{2} \, A_\mu^a \, m^2 A_\mu^a
+ i B^a \d_\mu A_\mu^a + \d_\mu \bar{c}^a \d_\mu c^a \right) \,.
\label{g0action}
\ee

Performing a Wilsonian renormalization group transformation on this free
field theory (see, e.g., ref.\ \cite{Bel91}), the only nontrivial step
is the rescaling of the space-time coordinates $x$ and the fields. Writing
the rescaling of the coordinates as $x \to x/s$ with a parameter $s > 1$,
the action \eqref{g0action} is invariant under the renormalization
group transformation if we rescale the gluon field as
\be
A_\mu^a (x) \to s^{D/2} A_\mu^a (s x) \,. \label{gluonscal}
\ee
This behavior under rescaling defines the canonical dimension of the gluon 
field which is hence $D/2$ and manifestly differs from its usual canonical
dimension $(D-2)/2$. The reason is that the action \eqref{g0action} 
corresponds to the high-temperature fixed point rather than the usually
considered critical Gaussian fixed point \cite{Bel91}, as
far as the gluon field is concerned. The high-temperature
fixed point is naturally related to the model of the stochastic vacuum. The
canonical dimension of the ghost fields, on the other hand, is the ordinary
$(D-2)/2$.

Reintroducing the coupling terms of Faddeev-Popov theory into the action,
the scaling dimensions around the infrared fixed point action \eqref{g0action}
turn out to be different for the different couplings. Concretely, both the 
three-gluon and the four-gluon couplings become perturbatively 
\emph{irrelevant}, while the ghost-gluon coupling is relevant only for
dimensions $D < 2$ (and marginal at $D=2$). In other words, the upper
critical dimension of the theory is two. This can also be concluded from
an infrared power counting analysis to all orders of perturbation theory
\cite{TW11}.

Hence, for an infrared analysis of the theory in dimensions $D > 2$, we
would be justified to work with the free action \eqref{g0action} alone,
which obviously directly leads to the decoupling solution for the
propagators (in the extreme infrared limit). It will turn out to be very
interesting, however, to include the coupling that is marginal at $D=2$
dimensions and to treat dimensions above two in an epsilon expansion.
We will hence add the coupling term
\be
S^1_{\text{IR}} = m \, \kappa^{(2-D)/2} \, \bar{g} f^{abc} \int d^D x \, 
\d_\mu \bar{c}^a A_\mu^b \, c^c \label{interact}
\ee
to the free action \eqref{g0action}, where we have introduced the
dimensionless coupling constant $\bar{g}$. Incidentally, keeping the
ghost-gluon coupling while neglecting the three- and four-gluon
couplings is precisely equivalent to keeping only the diagrams with the
biggest number of (internal) ghost propagators, that is, ghost dominance.

\subsection{Infrared fixed point}

In the theory defined by $S^0_{\text{IR}} + S^1_{\text{IR}}$, the one-loop
expressions for the propagators are precisely those represented in 
fig.\ \ref{IRDS}, except that the propagators in the loop diagrams should
be replaced with the tree-level propagators. We have given the result for
the ghost self-energy in $D = 2 + \epsilon$ dimensions already in eq.\
\eqref{ghostSE2}. For the gluon self-energy, we get in $D = 2 + \epsilon$ 
dimensions
\bal
\parbox{3.1cm}{\begin{center}
  \begin{tikzpicture}[>=stealth,scale=1.4]
   \begin{scope}[snake=coil, segment length=3pt, segment amplitude=2pt,%
   line before snake=1.5pt]
   \draw[snake] (-0.45,0) -- (0.05,0);
   \draw[snake] (0.98,0) -- (1.44,0) node[below] {$\scs p$};
   \end{scope}
   \draw[dash pattern=on 1.5pt off 1.2pt] (0.5,0) circle (0.5cm);
   \draw[->,very thin] (0.49,-0.5) -- (0.52,-0.5) node[below] {$\scs k-p$};
   \draw[->,very thin] (0.51,0.5) -- (0.48,0.5) node[above] {$\scs k$};
   \fill (0,0) circle (1.2pt);
   \fill (1,0) circle (1.2pt);
  \end{tikzpicture}
  \end{center}}
&= -N m^2 (\kappa^2)^{(2-D)/2} \, \bar{g}^2 \delta^{ab} 
\int \frac{d^D k}{(2 \pi)^D} \, 
(k_\mu - p_\mu) \frac{1}{k^2} \, k_\nu \, \frac{1}{(k - p)^2} \, 
\n \\[-6mm]
&= \frac{1}{2} \, \frac{N \bar{g}^2}{4 \pi} \, m^2 \delta^{ab}
\left[ \left( \frac{2}{\epsilon} + \gamma_E - \ln (4 \pi)
+ \ln \frac{p^2}{\kappa^2} - 2 \right) \delta_{\mu \nu} 
+ 2 \, \frac{p_\mu p_\nu}{p^2} \right] \,. \label{gluonSE}
\eal
It is important for the consistency of the renormalization procedure that
the counterterm can be chosen local, in this case proportional to $m^2
\delta_{\mu\nu}$. The gluon self-energy is \emph{not} transverse in 
momentum, which is not a surprise since BRST invariance is broken. Only the 
transverse part, given by the coefficient of $\delta_{\mu\nu}$ in the 
expression \eqref{gluonSE}, enters the result for the gluon propagator. 

For the renormalization of the theory, we introduce renormalized fields
as $A^a_\mu = Z_A^{1/2} A^a_{R,\mu}$ and $c^a = Z_c^{1/2} c^a_R$ and
normalize the renormalized propagators to
\bal
\big\langle A_{R,\rho}^a (p) A_{R,\sigma}^b (-q) 
\big\rangle \, \Big|_{p^2 = \mu^2}
&= \frac{1}{m^2} \left( \delta_{\rho \sigma} - \frac{p_\rho p_\sigma}{p^2} 
\right) \delta^{ab} (2 \pi)^D \delta (p - q) \,, \n \\
\big\langle c_R^a (p) \bar{c}_R^b (-q) \big\rangle \, \Big|_{p^2 = \mu^2}
&= \frac{1}{\mu^2} \, \delta^{ab} (2 \pi)^D \delta (p - q)
\label{normcond}
\eal
at the renormalization scale $\mu$. Using the expressions \eqref{ghostSE2}
and \eqref{gluonSE} for the self-energies, the normalization conditions
\eqref{normcond} define the field renormalization constants $Z_A$ and $Z_c$
as functions of the scale $\mu$. The one-loop anomalous dimensions are then 
easily calculated to be
\be
\gamma_A = \mu^2 \frac{d}{d \mu^2} \ln Z_A = \frac{1}{2} \, 
\frac{N \bar{g}^2}{4 \pi} \,,
\qquad \gamma_c = \mu^2 \frac{d}{d \mu^2} \ln Z_c = - \frac{1}{2} \, 
\frac{N \bar{g}^2}{4 \pi} \,. \label{anomdim1}
\ee

We define the dimensionless renormalized coupling constant as usual from
the renormalized vertex function at the symmetric point,
\bal
\left. \Gamma_{R, \bar{c} A c} (p, q, r) 
\right|_{p^2 = q^2 = r^2 =\mu^2} 
&= Z_A^{1/2} (\mu) Z_c (\mu) \left. \Gamma_{\bar{c} A c} (p, q, r) 
\right|_{p^2 = q^2 = r^2 =\mu^2} \n \\
&= m \, \mu^{-\epsilon/2} \, \bar{g}_R (\mu) f^{a b c} i p_\rho \, (2 \pi)^D
\delta(p + q + r) \label{gRdef}
\eal
(cf.\ eq.\ \eqref{interact}), where we have omitted the Lorentz and color 
indices on the symbol $\Gamma_{\bar{c} A c}$ of the vertex function. We
can now calculate the beta function. Taylor's non-renormalization theorem
implies that the bare vertex function at the symmetric point is
independent of the scale $\mu$, so that the one-loop beta function becomes
\be
\beta (\epsilon, \bar{g}_R) = \mu^2 \frac{d}{d \mu^2} \, \bar{g}_R 
= \frac{1}{2} \, \bar{g}_R \left( \frac{\epsilon}{2} + \gamma_A 
+ 2 \gamma_c \right)
= \frac{1}{2} \, \bar{g}_R \left( \frac{\epsilon}{2} - \frac{1}{2} \,
\frac{N \bar{g}_R^2}{4 \pi} \right) \,. \label{beta1}
\ee

From eq.\ \eqref{beta1} we identify two fixed points for
$\epsilon > 0$ ($D>2$): a trivial infrared-stable one, and an 
infrared-unstable one at
\be
\frac{N \bar{g}_R^2}{4 \pi} = \epsilon \,. \label{unstable1}
\ee
We begin with the discussion of the infrared-unstable fixed point. The 
$\mu$-independence of the bare propagators leads to differential 
equations for the renormalized propagators: using the definition of the 
anomalous dimensions we obtain the Callan-Symanzik equations
\bal
\mu^2 \frac{d}{d \mu^2} \, \big\langle A_{R,\rho}^a (p) 
A_{R,\sigma}^b (-q) \big\rangle
&= - \gamma_A \, \big\langle A_{R,\rho}^a (p) 
A_{R,\sigma}^b (-q) \big\rangle \,, \n \\
\mu^2 \frac{d}{d \mu^2} \, \big\langle c_R^a (p) 
\bar{c}_R^b (-q) \big\rangle
&= - \gamma_c \, \big\langle c_R^a (p) 
\bar{c}_R^b (-q) \big\rangle \,. \label{ren2ptdiff}
\eal
Inserting the fixed point value \eqref{unstable1} of the coupling constant 
in the anomalous dimensions, the differential equations for the renormalized 
propagators may be integrated with the normalization conditions 
\eqref{normcond} as initial conditions. The results are
\bal
\big\langle A_{R,\rho}^a (p) A_{R,\sigma}^b (-q) \big\rangle
&= \frac{1}{m^2} \left( \frac{p^2}{\mu^2} \right)^{\epsilon/2}
\left( \delta_{\rho \sigma} - \frac{p_\rho p_\sigma}{p^2} 
\right) \delta^{ab} (2 \pi)^D \delta (p - q) \,, \n \\
\big\langle c_R^a (p) \bar{c}_R^b (-q) \big\rangle 
&= \frac{1}{p^2} \left( \frac{\mu^2}{p^2} \right)^{\epsilon/2} 
\delta^{ab} (2 \pi)^D \delta (p - q) \,. 
\eal
This is \emph{exactly} the infrared behavior of the first scaling solution
\eqref{DSsol1}. So the renormalization group is able to reproduce one of
the scaling solutions in the infrared regime and furthermore demonstrates 
that it corresponds to an infrared-\emph{unstable} fixed point.

We should clarify that we refer to eqs.\ \eqref{ren2ptdiff} as (simple 
examples of) Callan-Symanzik equations for lack of a better name and
following the (rather recent) general use of the term, see, e.g., ref.\ 
\cite{Bel91}.
However, these are not precisely the equations originally derived
by Callan and Symanzik, but rather homogeneous renormalization group
equations that result from the reparameterization invariance of the bare
correlation functions.

Now we turn to the trivial infrared-stable fixed point. The fact that the 
stable fixed point is at $\bar{g}_R = 0$ implies that perturbation theory with 
the action $S^0_{\text{IR}} + S^1_{\text{IR}}$ should give a good description 
of the infrared regime. In fact, it was shown in refs.\ \cite{TW10,TW11}
that one-loop perturbation theory reproduces the lattice results very well 
in four dimensions and at least qualitatively in three dimensions. The
one-loop contributions that include the three- and four-gluon vertices
were taken into account in this comparison, however, they are suppressed
by powers of $(p^2/m^2)$ for small momenta.

Actually, the renormalization group approach can be used to go beyond
the mere fact that the decoupling solution in the extreme infrared limit
corresponds to a stable fixed point. The way the coupling constant
approaches the fixed point value (zero) can be obtained by integrating the
differential equation \eqref{beta1} for $\bar{g}_R (\mu)$. The result is
\be
\frac{N \bar{g}_R^2 (\mu)}{4 \pi} = \frac{(\mu^2/\Lambda^2)^{\epsilon/2}}
{1 + (\mu^2/\Lambda^2)^{\epsilon/2}} \, \epsilon \,, \label{decouprunning}
\ee
where we have introduced a new characteristic scale $\Lambda$. It is defined
as
\be
\frac{N \bar{g}_R^2 (\Lambda)}{4 \pi} = \frac{\epsilon}{2}
\ee
and should eventually be related to $\Lambda_{\text{QCD}}$ (this relation,
however, is beyond the scope of the present approach). Substituting the
result \eqref{decouprunning} for $\bar{g}$ in the anomalous dimensions
\eqref{anomdim1}, the integration of the differential equations
\eqref{ren2ptdiff} yields
\bal
\big\langle A_{R,\rho}^a (p) A_{R,\sigma}^b (-q) \big\rangle
&= \frac{1}{m^2} \, \frac{1 + (p^2/\Lambda^2)^{\epsilon/2}}
{1 + (\mu^2/\Lambda^2)^{\epsilon/2}} 
\left( \delta_{\rho \sigma} - \frac{p_\rho p_\sigma}{p^2} 
\right) \delta^{ab} (2 \pi)^D \delta (p - q) \,, \n \\
\big\langle c_R^a (p) \bar{c}_R^b (-q) \big\rangle 
&= \frac{1}{p^2} \, \frac{1 + (\mu^2/\Lambda^2)^{\epsilon/2}}
{1 + (p^2/\Lambda^2)^{\epsilon/2}} \, 
\delta^{ab} (2 \pi)^D \delta (p - q) \,. \label{decoupl}
\eal

The infrared behavior of the propagators predicted by eq.\ \eqref{decoupl}
is at least qualitatively confirmed by lattice calculations in $D=3$
dimensions \cite{CM09}. In $D=4$ space-time dimensions ($\epsilon = 2$), 
the increase of the 
gluon propagator proportional to $p^2$ starting from a non-zero constant at 
$p^2 = 0$ as described by eq.\ \eqref{decoupl} is actually seen \cite{CM10} 
in lattice calculations at lattice parameter $\beta = 0$, which in 
continuum language means that the proper Yang-Mills action 
\be
\int d^D x \, \frac{1}{4} F^a_{\mu\nu} F^a_{\mu\nu}
\ee
is omitted in the probability distribution, and only the delta functional
\eqref{deltaf} and the Faddeev-Popov determinant \eqref{fpdet} (and the 
restriction to the Gribov region) are left. 

In order to reproduce the
$D=4$ lattice results with the Yang-Mills action taken into account, we
have to reintroduce the term \eqref{cinetic} into the action $S^0_{\text{IR}} 
+ S^1_{\text{IR}}$. The reason is that this term, although perturbatively
irrelevant given the scaling behavior \eqref{gluonscal} of the gluon field,
is of the same order in $(p^2/m^2)$ as the correction to the tree-level
gluon propagator obtained from the renormalization group improvement in eq.\
\eqref{decoupl} for $\epsilon = 2$. Since at the one-loop level there are no 
quantum corrections to \eqref{cinetic} considered as a composite operator, 
we can just add this term to the 
renormalization-group improved result \eqref{decoupl} for the proper gluonic 
two-point function (inverse propagator for the transverse part), which
results in
\be
\big\langle A_{R,\rho}^a (p) A_{R,\sigma}^b (-q) \big\rangle
= \left( p^2 + m^2 \, \frac{1 + \mu^2/\Lambda^2}{1 + p^2/\Lambda^2} 
\right)^{-1} \left( \delta_{\rho \sigma} - \frac{p_\rho p_\sigma}{p^2} 
\right) \delta^{ab} (2 \pi)^D \delta (p - q) \,.
\ee
This is precisely of the form \eqref{gluonRGZ} of the infrared propagator 
found in the refined Gribov-Zwanziger framework (without an additional
$\langle A^a_\mu A^a_\mu \rangle$-condensate), with the effective 
momentum-dependent mass given by (cf.\ eqs.\ \eqref{gluonRGZ2}, 
\eqref{effmass})
\be
M^2_{\text{eff}} (p^2) = \frac{m^2 (\mu^2 + \Lambda^2)}{p^2 + \Lambda^2} \,.
\ee

A different renormalization scheme based on the full Faddeev-Popov action
\eqref{fadpop} plus the gluon mass term \eqref{mass} has been proposed in 
ref.\ \cite{TW11}. The corresponding Callan-Symanzik equations reproduce
the decoupling solutions as measured on the lattice reasonably well over the
whole momentum range in four dimensions, and at least qualitatively in three 
dimensions.

In the special case of two space-time dimensions ($\epsilon = 0$), the
two fixed points considered so far coalesce, and the result is one
infrared-unstable trivial fixed point. The instability implies that the
free action \eqref{g0action} cannot describe the infrared physics in two
dimensions. Indeed, calculations on the lattice confirm that in two
dimensions the decoupling solution corresponding to the action 
\eqref{g0action} is \emph{not} realized \cite{Maa07}. Instead, apparently
the second scaling solution \eqref{DSsol2} is found, and we have not yet
succeeded in giving a description of the corresponding fixed point 
(see, however, below).

\subsection{Lifshitz point}

Up to now, we have not made any attempt to implement the horizon condition 
in the form \eqref{infdivgh}. It is not difficult to do so: on the
classical level, for a local (and covariant) formulation, we have to
replace the ghost term in the action \eqref{g0action} with
\be
\int d^D x \, \frac{1}{b^2} \d_\mu \bar{c}^a (-\d^2) \d_\mu c^a \,,
\label{lifpoint}
\ee
where we have introduced an arbitrary dimensionful parameter $b$. The
form \eqref{lifpoint} of the action is known in statistical physics as
an isotropic Lifshitz point \cite{HLS75,MC98}.

Using the term \eqref{lifpoint} in the free action changes the 
canonical dimension of the ghost fields. Scale invariance of the
free action now implies the scaling behavior
\be
c^a (x) \to s^{(D-4)/2} \, c^a (s x) \,,
\qquad \bar{c}^a (x) \to s^{(D-4)/2} \, \bar{c}^a (s x) \,.
\label{ghostscal}
\ee
Then, around the Lifshitz point, the three- and four-gluon couplings
are perturbatively irrelevant just as before, while the
ghost-gluon coupling becomes relevant for dimensions $D < 6$.

In order to describe the infrared behavior of the theory with the 
horizon condition \eqref{infdivgh} implemented in dimensions
between two and four, we use an epsilon expansion around $D=6$.
We introduce the coupling term
\be
S^1_{\text{IR}} = \frac{m}{b^2} \, \kappa^{(6-D)/2} \, \bar{g} f^{abc} 
\int d^D x \, \d_\mu \bar{c}^a A_\mu^b \, c^c \label{interact2}
\ee
with the dimensionless coupling constant $\bar{g}$ (different
from the coupling constant in eq.\ \eqref{interact}),
and follow the procedure applied before in close analogy, 
starting with the calculation of
the ghost and gluon self-energies, now at dimension 
$D = 6 - \epsilon$ and with the tree-level ghost propagator
proportional to $b^2/k^4$. We stress that a fine tuning of the
tree-level ghost term that appears in eq.\ \eqref{g0action} is
necessary in order to fulfill the horizon condition \eqref{infdivgh} 
to one-loop order, just like in ordinary $\phi^4$-theory the 
tree-level mass term has to be fine-tuned to reach the 
critical point.

We use normalization conditions and a definition of the 
renormalized coupling constant $\bar{g}_R (\mu)$ in strict
analogy to eqs.\ \eqref{normcond} and \eqref{gRdef}, adapted to 
the new form \eqref{lifpoint} of the ghost term and the coupling 
\eqref{interact2}. Taylor's non-renormalization theorem continues
to hold in the present situation, and the result for the one-loop
beta function is
\be
\beta (\epsilon, \bar{g}_R) = \mu^2 \frac{d}{d \mu^2} \, \bar{g}_R 
= -\frac{1}{2} \, \bar{g}_R \left( \frac{\epsilon}{2} - \frac{1}{2} \,
\frac{N \bar{g}_R^2}{(4 \pi)^3} \right) \,. \label{beta2}
\ee

It thus turns out that the infrared-stable fixed point for dimensions
$D < 6$ ($\epsilon > 0$) is the nontrivial one. The final result for
the renormalization-group improved infrared propagators 
corresponding to the infrared-stable fixed point is
\bal
\big\langle A_{R,\rho}^a (p) A_{R,\sigma}^b (-q) \big\rangle
&= \frac{1}{m^2} \left( \frac{p^2}{\mu^2} \right)^{\epsilon/12}
\left( \delta_{\rho \sigma} - \frac{p_\rho p_\sigma}{p^2} 
\right) 
\delta^{ab} (2 \pi)^D \delta (p - q) \,, \n \\
\big\langle c_R^a (p) \bar{c}_R^b (-q) \big\rangle 
&= \frac{b^2}{p^4} \left( \frac{p^2}{\mu^2} \right)^{5\epsilon/24} 
\delta^{ab} (2 \pi)^D \delta (p - q) \,. \label{scaling2}
\eal

In the notation of eq.\ \eqref{powerans}, we obtain the following
values for the infrared exponents:
\be
\alpha_F (D) = \frac{5 D - 6}{24} \,,
\qquad \alpha_G (D) = - \frac{18 - D}{12} \,. \label{CSsol2}
\ee
The exponents fulfill the sum rule \eqref{sumrule}. The values \eqref{CSsol2}
should be compared to the (approximate) exponents of the second scaling 
solution \eqref{DSsol2},
\be
\alpha_F (D) = \frac{5 D - 5}{25} \,, 
\qquad \alpha_G (D) = - \frac{16 - D}{10} \,. \label{DSsol2b}
\ee
The two results coincide (exactly) at $D=6$, and in the range $2 \le D \le 4$
they differ the most at $D=2$ (where the approximation \eqref{DSsol2b} turns 
out to be exact): $\alpha_F = 1/6$ vs.\ $\alpha_F = 1/5$ and 
$\alpha_G = -8/6$ vs.\ $\alpha_G = -7/5$. However, compared to our
experience with the epsilon expansion in statistical physics, the
coincidence of the infrared exponents \eqref{CSsol2} and \eqref{DSsol2b}
is unexpectedly good, and we consider them to be approximations to the
same solution of the exact Dyson-Schwinger equations of the 
theory.

\subsection{Discussion}

We have reproduced the two scaling solutions and the decoupling solution
of the Dyson-Schwinger equations with the help of an analytical renormalization
group analysis. Only the second scaling solution \eqref{DSsol2} and the
decoupling solution correspond to infrared-stable fixed points, while the
first scaling solution \eqref{DSsol1} corresponds to an infrared-unstable
fixed point in dimensions $D$ between two and four.

In fact, the Lifshitz point which leads to the second scaling solution is
only stable if one implements the horizon condition \eqref{infdivgh} (in
addition to the gluon mass term which is a natural result of the breaking of
BRST invariance). As remarked after eq.\ \eqref{interact2}, the implementation 
of the horizon condition requires a fine tuning of the tree-level term
\be
\int d^D x \, \d_\mu \bar{c}^a \d_\mu c^a \,, \label{ghostkin}
\ee
a relevant coupling under the scaling behavior \eqref{ghostscal} of the ghost 
fields. Consequently, the nontrivial fixed point in eq.\ \eqref{beta2}
is infrared-\emph{unstable} unless one insists in implementing the horizon
condition.

We hence conclude that in dimensions $D > 2$ the physically realized solution
according to lattice simulations and the refined Gribov-Zwanziger approach
is the only one that corresponds to an
infrared-stable fixed point. In other words, the only input in the
Callan-Symanzik approach is the gluon mass term that is generated by the
breaking of BRST invariance, then the renormalization group flow 
automatically leads to the decoupling solution without further
assumptions. In two space-time dimensions, on the other hand, the trivial
fixed point corresponding to the decoupling solution becomes unstable,
and the lattice results point to the second scaling solution in this
case. However, according to the explanation above the fixed point corresponding
to this scaling solution should also be unstable. Indeed, calculating the 
one-loop corrections to \eqref{ghostkin} considered as a composite operator, 
the fixed point becomes even more strongly infrared-repulsive with decreasing 
dimension. Hence, unless there is some reason to implement the horizon 
condition precisely in two dimensions, we do not yet have a satisfactory 
understanding of the infrared behavior of the theory in two dimensions
within the Callan-Symanzik approach.

In conclusion, we have presented a new approach to the infrared regime of
Yang-Mills theory in Landau gauge that is based on a renormalization group
analysis with Callan-Symanzik equations in an epsilon expansion.
In contrast to the Dyson-Schwinger equations, the Callan-Symanzik equations
are analytical and inherently systematic as an approximation scheme.
Furthermore, they are technically relatively simple and economic as 
compared to the Gribov-Zwanziger approach with the additional auxiliary
fields necessary for a local formulation. This should be an important
advantage in the further elaboration of the theory, in particular the
inclusion of condensates (as in the refined Gribov-Zwanziger framework)
and dynamical quarks.

\ack
Support by CIC-UMSNH and Conacyt project CB-2009/131787 is gratefully 
acknowledged.


\section*{References}

\begin{thebibliography}{99}

\bibitem{Gri78}
Gribov V N 1978
\textit{Nucl. Phys.} B \textbf{139} 1

\bibitem{Zwa04}
Zwanziger D 2004
\textit{Phys. Rev.} D \textbf{69} 016002

\bibitem{Zwa89}
Zwanziger D 1989
\textit{Nucl. Phys.} B \textbf{323} 513

\bibitem{Zwa93}
Zwanziger D 1993
\textit{Nucl. Phys.} B \textbf{399} 477

\bibitem{Zwa02}
Zwanziger D 2002
\textit{Phys. Rev.} D \textbf{65} 094039

\bibitem{SHA97}
von Smekal L, Hauck A and Alkofer R 1997
\textit{Phys. Rev. Lett.} \textbf{79} 3591
 
\bibitem{SHA98}
von Smekal L, Hauck A and Alkofer R 1998
\textit{Annals Phys.} \textbf{267} 1

\bibitem{Tay71}
Taylor J C 1971
\textit{Nucl. Phys.} B \textbf{33} 436
%%CITATION = NUPHA,B33,436;%%

\bibitem{CMM04}
Cucchieri A, Mendes T and Mihara A 2004
\textit{JHEP} \textbf{12} 012

\bibitem{Dal11}
Dall'Olio P 2011
\textit{El R\'egimen Infrarrojo en Teor\'ias de Yang-Mills}
MSc thesis (Universidad Michoacana de San Nicol\'as de Hidalgo)

\bibitem{FA02}
Fischer C and Alkofer R 2002
\textit{Phys. Lett.} B \textbf{536} 177

\bibitem{FMP09}
Fischer C S, Maas A and Pawlowski J M 2009
\textit{Annals Phys.} \textbf{324} 2408

\bibitem{LS02}
Lerche C and von Smekal L 2002
\textit{Phys. Rev.} D \textbf{65} 125006

\bibitem{PLN04}
Pawlowski J M, Litim D F, Nedelko S and von Smekal L 2004
\textit{Phys. Rev. Lett.} \textbf{93} 152002

\bibitem{Zwa94}
Zwanziger D 1994
\textit{Nucl. Phys.} B \textbf{412} 657

\bibitem{KO79}
Kugo T and Ojima I 1979
\textit{Prog. Theor. Phys. Suppl.} \textbf{66} 1

\bibitem{Kug95}
Kugo T 1995
\textit{Preprint} arXiv:hep-th/9511033

\bibitem{DGS08}
Dudal D, Gracey J A, Sorella S P, Vandersickel N and Verschelde H 2008
\textit{Phys. Rev.} D \textbf{78} 065047

\bibitem{AN04}
Aguilar A C and Natale A A 2004
\textit{JHEP} \textbf{0408} 057

\bibitem{BBL06}
Boucaud P, Br\"untjen T, Leroy J P, Le Yaouanc A,
Lokhov A, Micheli J, P\`ene O and Rodr\'{\i}guez-Quintero J 2006
\textit{JHEP} \textbf{0606} 001

\bibitem{Fra08}
Frasca M 2008
\textit{Phys. Lett.} B \textbf{670} 73

\bibitem{BIM07}
Bogolubsky I L, Ilgenfritz E M, M\"uller-Preussker M and Sternbeck A 2007
\textit{Proc. Sci.} LAT2007 290

\bibitem{CM07}
Cucchieri A and Mendes T 2007
\textit{Proc. Sci.} LAT2007 297

\bibitem{SSL07}
Sternbeck A, von Smekal L, Leinweber D B and Williams A G 2007
\textit{Proc. Sci.} LAT2007 340

\bibitem{Maa07}
Maas A 2007
\textit{Phys. Rev.} D \textbf{75} 116004

\bibitem{DSV11}
Dudal D, Sorella S P and Vandersickel N 2011
\textit{Phys. Rev.} D \textbf{84} 065039 

\bibitem{DOV10}
Dudal D, Oliveira O and Vandersickel N 2010
\textit{Phys. Rev.} D \textbf{81} 074505

\bibitem{CDM11}
Cucchieri A, Dudal D, Mendes T and Vandersickel N 2011
\textit{Preprint} arXiv:1111.2327 [hep-lat]

\bibitem{Zwa91}
Zwanziger D 1991
\textit{Nucl. Phys.} B \textbf{364} 127

\bibitem{CMT05}
Cucchieri A, Mendes T and Taurines A R 2005
\textit{Phys. Rev.} D \textbf{71} 051902

\bibitem{Web11}
Weber A 2011
%Epsilon expansion for infrared Yang-Mills theory in Landau gauge
\textit{Preprint} arXiv:1112.1157 [hep-th]

\bibitem{CF76}
Curci G and Ferrari R 1976
\textit{Nuovo Cim.} A \textbf{32} 151

\bibitem{TW10}
Tissier M and Wschebor N 2010
\textit{Phys. Rev.} D \textbf{82} 101701

\bibitem{TW11}
Tissier M and Wschebor N 2011
\textit{Phys. Rev.} D \textbf{84} 045018

\bibitem{Bel91} 
Le Bellac M 1991 
\textit{Quantum and Statistical Field Theory}
(Oxford: Oxford University Press)

\bibitem{CM09} 
Cucchieri A and Mendes T 2009
\textit{Proc. Sci.} QCD-TNT09 026.
%[arXiv:1001.2584]

\bibitem{CM10}
Cucchieri A and Mendes T 2010
\textit{Phys. Rev.} D \textbf{81} 016005

\bibitem{HLS75}
Hornreich R M, Luban M, and Shtrikman S 1975
\textit{Phys. Rev. Lett.} \textbf{35} 1678

\bibitem{MC98}
Mergulh\~{a}o, Jr. C and Carneiro C E I 1998
\textit{Phys. Rev.} B \textbf{58} 6047

\end{thebibliography}
\end{document}